\documentclass[cameraready]{Interspeech}

\title{PrefSQA: Pairwise Preference Prediction for Speech Quality Assessment and the Critical Role of High Quality Datasets}

\author[affiliation={1}, orcid=0009-0000-9911-6182, correspondingauthor]{Junyi}{Fan}
\author[affiliation={1}, orcid=0000-0002-7997-5403]{Donald S.}{Williamson}

\address{
    $^1$ Department of Computer Science and Engineering, The Ohio State University, USA
}

\email{fan.1188@osu.edu, williamson.413@osu.edu}


\usepackage{comment}

\usepackage{multirow}

\begin{document}

\maketitle

\begin{abstract}
    Mean opinion scores (MOS) are widely used for speech quality assessment, yet scalar labels are sensitive to rater variability and listening test differences. This introduces labeling noise, which limits the reliability of MOS prediction. Preference prediction reduces this variability as listeners compare signals directly, producing cleaner labels. We study MOS-free preference prediction and propose PrefSQA, which incorporates uncertainty-aware logits, an impairment attention head, and a module based on non-matching-reference comparisons. We use and refine five datasets, including MOS-derived and low-noise simulated sets with matching and non-matching content, experiment with human preference sets, and test on unseen data. Experiments show small improvements on MOS-derived data, while other sets reveal clear improvement over the baselines, highlighting the value of high-quality preference data and demonstrating the effectiveness of the proposed method.
\end{abstract}

\section{Introduction}

Perceptual speech quality assessment (SQA) plays a crucial role in speech enhancement, text-to-speech (TTS), and automatic speech recognition systems \cite{Loizou2011}. Although subjective listening tests remain the most reliable way to assess speech quality, they are expensive, time-consuming, and impractical to run at large scales required by modern speech systems. Automatic methods are therefore developed to efficiently estimate quality. Modern automatic SQA has long relied on mean opinion scores (MOS) to assign scalar values to utterances. Although widely adopted, MOS-based systems introduce challenges that limit progress in automatic SQA. Discrepancies in listening test protocols, variability among listeners \cite{ZequeiraJimenez2022InfluencingFactorsSQA}, and the use of discrete rating scales \cite{koster2015comparison} all contribute to high labeling noise \cite{huang2024mos}. The noise creates difficulty for supervised learning and obscures quality differences between signals. These issues motivated a shift toward preference-based assessment, where listeners compare quality levels of two signals rather than assign absolute scores. Pairwise judgments reduce subjective variability, as relative quality is easier and more consistent for listeners \cite{10.1371/journal.pone.0190393, phelps2015pairwise}. As a result, the judgments also depend less on listening test protocols and their results translate well across different tests.

Deep learning has enabled rapid developments in modern SQA systems. MOSNet \cite{lo2019mosnet}, NISQA \cite{mittag2021nisqa}, and VoiceMOS Challenges \cite{huang2024voicemos}, to name a few, demonstrated substantial progress over traditional metrics. However, corpus and domain effect issues \cite{huang2024mos} limit performance for these MOS prediction models. To overcome this, preference‑based ideas have been incorporated in various ways, such as adding pairwise‑comparison losses as auxiliary supervision to refine MOS regression \cite{wang2023mospc}, introducing preference‑ or ranking‑based objectives to regularize quality representations \cite{serra2021sesqa, manocha2021noresqa}, training no‑reference models from pairwise preferences to reduce reliance on noisy absolute labels \cite{manocha2022sqapp}, or incorporating comparative MOS or pairwise objectives as components of a unified training recipe \cite{wang2026urgentmos}. However, these approaches, along with a related study on predicting pairwise preferences between TTS stimuli \cite{valentini2022predicting}, remain tied to MOS or other rating-derived supervision, as they are incorporated into the pipelines at one or more stages. This has resulted in a limited number of existing works focusing on MOS‑independent preference prediction, with \cite{Shi2025UPPSQA} being, to the best of our knowledge, the only major work addressing this issue. Meanwhile, public datasets for preference-based SQA are scarce compared to MOS datasets. This forces preference studies to often rely on labels derived from MOS, which were designed for scalar rating tasks. MOS and preference listening tests use substantially different setups, causing the derived datasets not only to bear the inherent noise from MOS data but also to introduce additional noise due to discrepancies in study protocols. This can obscure major improvements in supervised preference prediction, where accuracy differences of multiple models are smaller than actual differences in model capabilities.

This work studies MOS-free pairwise preference prediction through PrefSQA and examines how dataset quality affects the visibility of model improvements. We construct five preference datasets based on SOMOS \cite{maniati22_interspeech}, NISQA \cite{mittag2021nisqa}, LibriSpeech \cite{panayotov2015librispeech}, and ChiME-3 \cite{barker2015third}, which include pairs with matching and non-matching lexical content, with both real and simulated utterances. Despite being derived from existing MOS datasets, MOS labels are never seen by models in any way during development. This reduces the labeling noise they introduce. More importantly, this also aligns better with real-world situations where preference labels are not always available with MOS. Building on the framework of UPPSQA \cite{Shi2025UPPSQA}, we adopt its semantic-acoustic architecture, which combines wav2vec 2.0 \cite{baevski2020wav2vec} and WavLM \cite{chen2022wavlm}, as the starting point for our model design. We then introduce original architectural decisions to improve pairwise preference modeling. Specifically, our model is augmented with uncertainty-aware Bradley-Terry preference logits \cite{bradley1952rank}, a lightweight impairment attention head that emphasizes local degradations, and a feature-level non-matching-reference (NMR) head that refines global rankings through in-batch comparisons. Results on MOS-derived datasets show small gains, while results on high-quality simulated datasets reveal clear performance gaps across models, showing how low-noise labels are essential for identifying benefits of architectural changes. Evaluations on human-preference labels and unseen data further confirm the effectiveness of our model.

\begin{figure*}[t]
  \centering
  \includegraphics[width=0.85\textwidth]{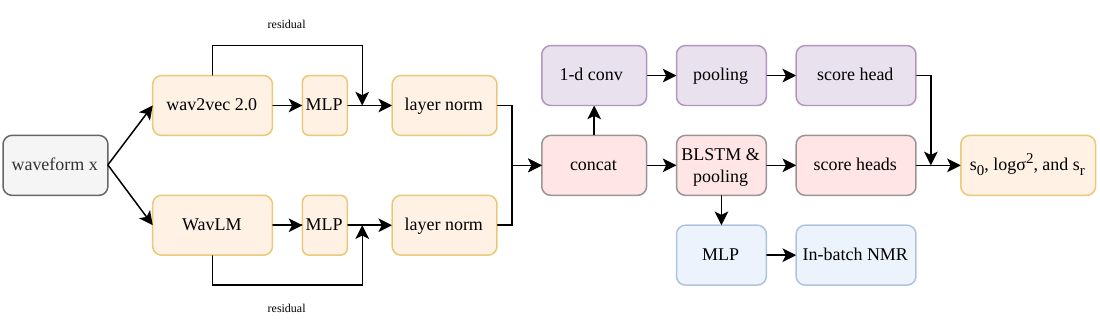}
  \caption{PrefSQA model architecture for input waveform $x$ with semantic-acoustic encoders, augmented with uncertainty-aware preference logits, a lightweight impairment attention head (purple blocks), and a feature-level non-matching-reference (NMR) head (blue blocks). The other input waveform $y$ (not shown here) goes through the same process.}
  \label{fig:Prefsqa_arch}
\end{figure*}

\section{Method}
\label{sec:method}

\subsection{Backbones: dual encoders}

Figure \ref{fig:Prefsqa_arch} illustrates the PrefSQA architecture for a single signal in a pair. Following \cite{Shi2025UPPSQA}, pretrained wav2vec2 and WavLM encoders provide semantic and acoustically sensitive representations, respectively, for a waveform. The wav2vec2 branch uses the last hidden state, while the WavLM branch passes the full set of layer outputs through a learnable layer-weighted sum module that produces a single sequence of hidden states. The layer-weighted sum module maintains a learnable parameter vector over layers and a temperature parameter that controls the sharpness of softmax weights. During training, a small gate randomly drops some layers before renormalization to prevent over-reliance on a small subset of WavLM layers and to encourage robust use of the encoder. The semantic and acoustic sequences then pass through a residual feature processor of two linear layers with GELU activation and a layer normalization on the residual sum of input and output. This preserves the original dimensionality while allowing task-specific adaptation. The processed semantic and acoustic features are concatenated along the channel axis and fed into a BLSTM. The BLSTM output sequence is pooled by averaging across time, yielding a vector embedding per utterance that summarizes both semantic and acoustic cues. Two linear heads map this embedding to a scalar score $s_0$ and a scalar log variance $\log \sigma^2$. $s_0$ behaves as a latent MOS value in an arbitrary scale, while the variance provides an uncertainty estimate that later controls the sharpness of the preference comparison between two utterances.

\subsection{Impairment attention head}
To emphasize local impairment cues, PrefSQA employs an impairment attention head that operates on encoder features. This branch applies a 1-D convolution over time on the concatenated encoder features, followed by a sigmoid gating layer that produces a time-wise attention mask. This forms an attention-weighted average of the input that later passes through a multilayer projection, offering a scalar residual $s_r$. This residual, scaled by a factor $\alpha$, is added to $s_0$ and acts as a lightweight distortion-focused correction besides the global BLSTM summary. The final quality score is $s = s_0 + \alpha * s_r$.

\subsection{Uncertainty-aware pairwise preference logits}

Given utterance pair $x$ and $y$, the model computes their scores and log variances, $(s_x, \log \sigma_x^2)$ and $(s_y, \log \sigma_y^2)$. The preference logit for ``$x$ preferred over $y$'' is obtained as

\begin{align}
z_{x,y} = \frac{s_x - s_y}{\sqrt{\sigma_x^{2} + \sigma_y^{2} + \epsilon}}, 
\end{align}
where the denominator, denoted as $\tau$, acts as an uncertainty-dependent temperature. To avoid excessively sharp or overly flat preference logits, we clamp $\tau$ to a fixed range $[a,b]$ by $\min(\max(\tau, a), b)$. The final preference probability for $x$ over $y$ is the sigmoid of $z_{x,y}$. For training, however, we optimize the raw logits directly using binary cross-entropy (BCE) with logits, which applies sigmoid internally for numerical stability (Section~\ref{section:training objective}). Because the model only uses model-predicted scores and variances at the utterance level, it is valid regardless of whether the two signals share content or differ in length.

\subsection{Lightweight feature-level NMR preference head}

To further improve the global ranking, we include a feature-level NMR head \cite{manocha2021noresqa}. It operates on embeddings from a batch and uses pseudo-labels derived from the model's own scores. Specifically, after a BLSTM and average pooling, an MLP produces the utterance embedding $\mathbf{f}$, which is later fed into the NMR head. Given a batch of $B$ preference pairs, we form a bag of $2B$ utterance items by first collecting utterances from all pairs and retrieving their embeddings $\mathbf{f}$. Each item is then used once as an anchor, and for each anchor, up to $k$ distinct partner items are sampled without replacement from the remaining $2B-1$ items. For each pair, we construct a feature vector

\begin{align}
\mathbf{u}_{i,j} = [\mathbf{f}_i, \mathbf{f}_j, \mathbf{f}_i - \mathbf{f}_j, |\mathbf{f}_i - \mathbf{f}_j|].
\end{align}
where $\mathbf{f}_i$ and $\mathbf{f}_j$ are embeddings of the utterances. An MLP maps $\mathbf{u}_{i,j}$ to a scalar logit $\ell_{i,j}$ that predicts the probability that $i$ has higher quality than $j$. The corresponding soft target $t_{i,j}$ is the sigmoid of $(s_i - s_j)/\tau_{\text{nmr}}$, built from the latent scores with a fixed temperature $\tau_{\text{nmr}}$ and light label smoothing toward 0.5. An NMR loss, $\mathcal{L}_{\text{NMR}}$, is then defined as the BCE between $\ell_{i,j}$ and $t_{i,j}$, averaged over all sampled in-batch pairs. This auxiliary head encourages the embedding space to respect the order implied by the model's own scores in a more fine-grained way than the main pairwise loss alone.

\subsection{Training objectives}
\label{section:training objective}

The primary loss $\mathcal{L}_{\text{BT}}$ in \eqref{eq:bt_loss} is the Bradley-Terry logistic loss \cite{bradley1952rank} on $z_{x,y}$, where $m$ indexes pairs in a batch and $c[m]$ equals 1 if the label indicates $x$ is preferred and 0 otherwise. The NMR loss $\mathcal{L}_{\text{NMR}}$ is scaled by $\lambda$ and then added to the primary loss to form the total loss $\mathcal{L}$ in \eqref{eq:total_loss}.
\begin{gather}
\mathcal{L}_{\text{BT}} = \operatorname{BCEwithLogits}\bigl(z_{x,y}[m], c[m]\bigr) \label{eq:bt_loss} \\
\mathcal{L} = \mathcal{L}_{\text{BT}} + \lambda \mathcal{L}_{\text{NMR}} \label{eq:total_loss}
\end{gather}

\section{Datasets}
\label{sec:datasets}

\subsection{MOS-derived datasets: NISQA and SOMOS}
For the NISQA data, we use the full dataset including all subset conditions (e.g., P501, LIVE, SIM). We maintain all these subset conditions and also the original train, validation, and test split conditions. All pairs are generated inside those specific conditions and no utterances are moved across conditions. Within each condition, we randomly sample pairs made of utterances with non-matching content. Each utterance can be used at most three times. No pair is used more than once. The utterance with the higher MOS in a pair is labeled as preferred. Tie pairs with equal MOS are dropped, as we are not predicting ties. We then concatenate all train, validation, and test pairs across all subset conditions. Similar to NISQA, we follow the original split in SOMOS, made of TTS samples. Only SOMOS-clean is used. We then sample pairs from utterances with the same underlying sentence text for matching-content pairs or with different sentence texts for non-matching-content pairs. For both NISQA and SOMOS, the preferred signal appears as the first or second one in the pair randomly.

\subsection{CHiLi matching and non-matching datasets}
To obtain preference labels with minimal labeling noise, we construct pairwise matching (M) and non-matching (NM) datasets, named CHiLi, by mixing clean speech from LibriSpeech with additive background noise from CHiME-3. Both datasets share the same generation pipeline and differ only in whether two signals in a pair use the same clean speech or not. For the pairs, we define three scalar parameters: a base SNR $n_a$ for signal A sampled uniformly from -20 to 30 dB, an SNR difference $\Delta$ sampled uniformly from 0.5 to 10 dB, and a random sign of either $+, -$ to determine the SNR $n_b$ for signal B, where $n_b = n_a \pm \Delta$. This ensures that absolute SNR values are within the range normally seen in the literature and SNR differences between pairs make the task reasonably challenging. The SNR values for the resulting mixture are achieved by scaling the noise segment. The preferred signal in a pair is defined as the one with higher SNR. In the NM case, each sample appears exactly twice in two different pairs to achieve balanced representation and equal-sized datasets between M and NM. For any clean speech, the noise signal is always randomly selected from all conditions in CHiME-3. Thus, the background noise for two clean samples in any pair are essentially always different and sometimes even from different noise conditions, which greatly increases data diversity and therefore benefits model robustness.

\section{Experimental Results}
\label{sec:results}

\subsection{Experimental setup}
Input signals are resampled to 16 kHz, truncated or zero-padded to a maximum length of 6 seconds, with an attention mask marking valid samples. The layer-weighted sum module uses temperature 0.5 and gate dropout 0.1. The two-layer feature processors for the encoders each have a 64-dimensional bottleneck. A single-layer BLSTM has 256 hidden units per direction, followed by an MLP with hidden sizes 256-128-64. The impairment attention head has scaling factor $\alpha$ of 0.1 and uses a temporal convolution with 128 channels and kernel size 5, followed by a 1 by 1 gate convolution. $\tau$ for $z_{x,y}$ is clamped to $[0.6, 2.0]$. The NMR loss samples $k = 3$ partners per anchor, with temperature 1.3, label smoothing 0.03, and $\lambda$ of 0.9, using a two-layer MLP of size 256-128-1. Training uses AdamW \cite{loshchilov2019adamw}, with an effective batch size of 32, learning rate 0.001 for the task heads and wav2vec2, base learning rate 0.00003 for WavLM with layerwise decay factor 0.95, weight decay 0.01, and gradient norm clipping at 1.0. WavLM uses a smaller learning rate to retain stable pretrained acoustic features.

\subsection{Dataset setup}
All datasets in Section~\ref{sec:datasets} have balanced preference labels both globally and locally. Specifically, we partition pairs by absolute MOS or SNR difference into bins of width 0.1 and verify that in bins with enough samples, the two labels occur in approximately equal numbers. Ensuring balanced datasets is important for this task as it removes trivial strategies such as always predicting the majority class, which allows accuracy to reflect how well the model performs rather than label frequency bias. We also use SpeechEval \cite{wang2025speechllm} (tie pairs excluded) and SpeechJudge \cite{zhang2025speechjudge} to reflect performance on labels collected from real listening tests. It is worth mentioning that these two datasets only contain pairs with matching speech content. We also use the IUB dataset \cite{dong2020pyramid}, constructed from the COSINE corpus through MUSHRA tests, solely for testing to assess the model's generalization capabilities. We use its scaled MOS for all three audio samples (reference, anchor, and test) in a MUSHRA test to derive three audio pairs. Preference labels derived this way can be roughly treated as authentic labels due to how MUSHRA tests are conducted. One COSINE test set is constructed with randomly selected MUSHRA tests from the original dataset.

\subsection{Preference accuracy results}

Table~\ref{tab:accuracy} presents experimental results for SQAPP \cite{manocha2022sqapp}, UPPSQA \cite{Shi2025UPPSQA}, PrefSQA Frozen, and PrefSQA, trained and evaluated separately on selected datasets with their pair counts. It also reports results on the unseen IUB-COSINE test set using the same checkpoints trained on CHiLi NM or SpeechEval, rendering two different combinations. UPPSQA and the first stage of SQAPP are the two most closely related models that can be applied directly to our task. We implement both based on the original papers, as public code is not available. The difference between PrefSQA Frozen and PrefSQA is that dual encoders are frozen for the former during training. For NISQA and SOMOS data, PrefSQA models achieve the best results, with all models except SQAPP showing small differences. For SOMOS M and NM, the same three models offer close accuracies, despite the NM task being intuitively harder. The contrast with ChiLi datasets is much clearer, where our models dominate the performance by a large margin, with CHiLi M offering higher accuracies than NM across all four models. Our model also shows superior performance on SpeechEval and SpeechJudge, except that SQAPP performs the best on SpeechJudge. We assume this is because SpeechJudge provides direct, curated preference supervision that closely matches SQAPP’s training objective. We also achieve leading results with unseen data, as suggested by the IUB-COSINE set, tested on both CHiLi NM and Speecheval training checkpoints, which indicates strong generalization performance. 

\begin{table}[t]
  \caption{Prediction accuracy (\%) and dataset pair counts. Best and second-best scores are shown in bold and underlined, respectively. PrefSQA-F has the pretrained encoders frozen. C and S suffixes indicate results from checkpoints trained on CHiLi NM (\underline{N}on-\underline{M}atching) and SpeechEval, respectively.}
  \label{tab:accuracy}
  \centering
  \scriptsize
  \setlength{\tabcolsep}{1pt}
  \renewcommand{\arraystretch}{0.6}
  \begin{tabular}{l c c c c | c c c}
    \toprule
    \textbf{Dataset} &
    \textbf{SQAPP} & \textbf{UPPSQA} & \textbf{PrefSQA-F} & \textbf{PrefSQA} &
    \textbf{Train} & \textbf{Val} & \textbf{Test} \\
    \midrule
    NISQA$^{\dagger}$      & 64.83 & \underline{83.46} & 82.80 & \textbf{83.84} & 15917 & 3870 & 1052 \\
    SOMOS M$^{\dagger}$    & 65.54 & 71.96 & \textbf{73.27} & \underline{73.18} & 18055 & 1776 & 1721 \\
    SOMOS NM$^{\dagger}$   & 49.28 & 73.10 & \underline{73.48} & \textbf{74.72} & 20862 & 4450 & 4430 \\
    \midrule
    CHiLi M$^{\ddagger}$   & \underline{94.78} & 85.88 & 91.52 & \textbf{96.29} & 22831 & 2853 & 2855 \\
    CHiLi NM$^{\ddagger}$  & 86.90 & 81.05 & \underline{87.50} & \textbf{90.37} & 22831 & 2853 & 2855 \\
    \midrule
    SpeechEval$^{\S}$      & 80.24 & \underline{86.31} & \textbf{86.85} & 84.32 & 15443 & 3294 & 3163 \\
    SpeechJudge$^{\S}$     & \textbf{70.40} & 61.40 & 65.20 & \underline{68.20} & 42097 & 1000 & 1000 \\
    \midrule
    IUB-COSINE-C$^{\P}$       & 71.89 & \underline{78.06} & 77.22 & \textbf{83.50} & N/A & N/A & 1800 \\
    IUB-COSINE-S$^{\P}$       & 48.56 & 87.89 & \textbf{91.61} & \underline{89.28} & N/A & N/A & 1800 \\
    \bottomrule
  \end{tabular}

  \vspace{1mm}
  {\scriptsize $^{\dagger}$ MOS-derived \quad
   $^{\ddagger}$ simulated \quad
   $^{\S}$ human preference \quad
   $^{\P}$ unseen test-only}
\end{table}

\subsection{Dataset quality and error analysis}
\label{section:ccc}

Table~\ref{tab:margin_stats_corr}  (left) shows the concordance correlation coefficient (CCC) \cite{lawrence1989concordance} between ordered lists of misclassified pairs. Unlike Pearson correlation coefficient (PCC), a high CCC is a stronger indication that two models tend to make mistakes on similar pairs, not only in aggregate but also in the order induced by margin size, because it penalizes discrepancies in both scaling and mean level in addition to imperfect correlation. We first collect all misclassified test pairs, order them by their absolute MOS or SNR difference, and compute CCC between the margin lists of any two of the three selected models. We do this on all training sets except SpeechEval and SpeechJudge, which have no directly measurable statistics such as MOS or SNR. On MOS-derived sets, all CCC values are close to 1, indicating that the models make similar mistakes. Combined with small accuracy gaps in Table~\ref{tab:accuracy}, this suggests that labeling noise in MOS-derived labels affects performance visibility, especially in small margin regions highlighted by the error analysis in the next paragraph. In contrast, simulated ChiLi sets show much lower CCC, which suggests that our model can genuinely improve performance.

We also report the statistics of test pairs that PrefSQA misclassifies on all sets except SpeechEval and SpeechJudge. Given the same ordered lists obtained above, we group values in the lists into bins of width 0.1. Within each bin we divide the number of errors by the total number of samples in that bin, so that each bin contributes a normalized error rate rather than a raw count, removing the bias caused by some margin ranges containing more pairs than others. We summarize the distribution of bin centers through percentiles in Table~\ref{tab:margin_stats_corr} (right). Across the sets, the statistics confirm that errors concentrate in regions with small MOS or SNR margins, as values of the 50th percentile (P50), roughly representing the MOS or SNR ranges where the first half of misclassifications happen, are all much smaller than the values of P99-P50, representing the second half. This makes MOS-derived sets more vulnerable, as these regions happen to be those with dominating labeling noise.

\begin{table}[th]
  \caption{CCC between ordered lists of misclassified pairs on selected models and normalized error distribution (indicated by percentiles P50, P75, etc.) for PrefSQA as a function of absolute MOS or SNR margin.}
  \label{tab:margin_stats_corr}
  \centering
  \scriptsize
  \setlength{\tabcolsep}{2.5pt}
  \renewcommand{\arraystretch}{0.6}
  \begin{tabular}{l c c c | c c c c c | c}
    \toprule
    \textbf{Dataset} &
    \textbf{U-PF} & \textbf{U-P} & \textbf{PF-P} &
    \textbf{P50} & \textbf{P75} & \textbf{P90} & \textbf{P95} & \textbf{P99} &
    \textbf{P99-P50} \\
    \midrule
    NISQA$^{\dagger}$     & 0.94 & 0.97 & 0.90 & 0.43 & 0.68 & 1.12 & 1.33 & 2.23 & 1.80 \\
    SOMOS M$^{\dagger}$   & 0.99 & 0.93 & 0.92 & 0.33 & 0.53 & 0.78 & 1.08 & 1.48 & 1.15 \\
    SOMOS NM$^{\dagger}$  & 0.95 & 0.97 & 0.99 & 0.33 & 0.58 & 0.88 & 1.08 & 1.52 & 1.19 \\
    \midrule
    CHiLi M$^{\ddagger}$  & 0.56 & 0.30 & 0.51 & 1.38 & 2.23 & 3.23 & 3.68 & 4.43 & 3.05 \\
    CHiLi NM$^{\ddagger}$ & 0.80 & 0.61 & 0.78 & 2.08 & 3.78 & 5.43 & 7.47 & 9.18 & 7.10 \\
    \bottomrule
  \end{tabular}

  \vspace{1mm}
  {\scriptsize U-PF:  UPP vs Pref Frozen\quad
   U-P: UPP vs Pref \quad
   PF-P: Pref Frozen vs Pref}
\end{table}

\begin{table}[th]
  \caption{Ablation study accuracy (\%) on PrefSQA.}
  \label{tab:ablation}
  \centering
  \scriptsize
  \setlength{\tabcolsep}{4pt}
  \begin{tabular}{l c c c c}
    \toprule
    \textbf{Dataset} & \textbf{No Attn/NMR} & \textbf{No Attn} & \textbf{No NMR} & \textbf{Full} \\
    \midrule
    ChiLi M   & 95.52 & 96.22 & 96.25 & \textbf{96.29} \\
    ChiLi NM  & 89.35 & 88.86 & 90.12 & \textbf{90.37} \\
    \bottomrule
  \end{tabular}

\vspace{1mm}
  {\scriptsize Attn: impairment attention head\quad
   NMR: non-matching-reference head}
\end{table}

\subsection{Ablation study}

In the ablation study shown in Table~\ref{tab:ablation}, we compare variants that remove key components of PrefSQA on ChiLi. All accuracies are relatively high on ChiLi M, with the Full model being the highest. This pattern suggests that performance on easy data saturates even with components missing and that it becomes hard to attribute gains to individual modules. In contrast, ChiLi NM remains discriminative: removing the impairment attention head drops accuracy to 88.86 and removing the NMR head yields 90.12, both below the Full model. Although these gains are modest, the consistent improvement of the Full model suggests these two components provide complementary benefits when the task is harder.

\subsection{Discussion}
The accuracy results for our model do not consistently favor end-to-end fine-tuning, where the frozen and fine-tuned variants alternate as the best model across many datasets. This suggests that the benefit of adapting pretrained encoders depends on each preference dataset, where fine-tuning can improve task-specific representations on better-matched data, but can also underperform otherwise. This motivates a more controlled fine-tuning recipe that trades off stability and performance, for example by using L2SP regularization to determine how much encoder adaptation is optimal. We also notice that our fine-tuned model reaches peak validation accuracy with much fewer optimization steps than competing baselines on most datasets, while the frozen variant often converges even more rapidly.

\section{Conclusion and Future Work}
\label{sec:conclusion}

This paper studies MOS-free pairwise preference prediction for perceptual speech quality assessment by introducing PrefSQA, a dual encoder system that fuses wav2vec 2.0 and WavLM features with uncertainty-aware logits, impairment attention, and a lightweight in-batch NMR head to refine global rankings. The experiments show that MOS-derived datasets can obscure architectural advantages driven by labeling noise, while high-quality simulated datasets expose clearer performance gaps and highlight the benefits of the proposed architecture. Results also showcase satisfactory performance on real human preference data and unseen tests. A key limitation suggested by error analysis is that difficult cases concentrate in small margin regions, where two samples may plausibly be perceived as having the same quality. Future work to incorporate an explicit tie option for near indistinguishable pairs could potentially both improve accuracy and better align model outputs with human perception.

\section{Acknowledgment}
\label{sec:acknowledgment}
This work was supported in part by the Ohio Supercomputer Center, NSF award IIS-2235228, and NSF award IIS-2523648.

\section{Use of Generative AI Disclosure}
\label{sec:disclosure}
Generative AI tools have been used for editing and polishing this manuscript.

\bibliographystyle{IEEEtran}
\bibliography{mybib}

\end{document}